\documentclass{aa}
\usepackage{txfonts}
\usepackage{graphicx}
\usepackage{natbib}
\bibpunct{(}{)}{;}{a}{}{,}

\def\ms{m\,s$^{-1}$}
\def\fe{\ion{Fe}{i}\,6252}

\def\alfhor{$\alpha$~Hor}
\def\bethyi{$\beta$~Hyi}
\def\taucet{$\tau$~Cet}
\def\gamser{$\gamma$~Ser}
\def\betaql{$\beta$~Aql}
\def\muara{$\mu$~Ara}
\def\delpav{$\delta$~Pav}
\def\alfmen{$\alpha$~Men}

\begin{document}
\title{Bisectors of the cross-correlation function applied to stellar spectra\thanks{Based on observations collected at the 
    La Silla Observatory, ESO (Chile), with the HARPS
    spectrograph at the ESO 3.6m telescope (ESO Programme 075.D-0760), and on observations 
    obtained from the ESO/ST-ECF Science Archive Facility (ESO Programmes  073.D-0578, 073.C-0784 and 074.D-0380).}}
\subtitle{Discriminating stellar activity, oscillations and planets}

\author{T. H. Dall \inst{1}
  \and
  N. C. Santos \inst{2,3}
  \and
  T. Arentoft \inst{4}
  \and
  T. R. Bedding \inst{5}
  \and
  H. Kjeldsen \inst{4}
}

\offprints{T. H. Dall, \email{tdall@eso.org}}

\institute{European Southern Observatory, Casilla 19001, Santiago 19, Chile
  \and
  Centro de Astronomia e Astrof\'isica da Universidade de Lisboa, Observat\'orio Astron\'omico de Lisboa,
        Tapada da Ajuda, 1349-018 Lisboa, Portugal
  \and
  Observatoire de Gen\`eve, 51 Ch. des Maillettes, 1290 Sauverny, Switzerland
  \and
  Department of Physics and Astronomy, University of Aarhus, 8000 Aarhus C, Denmark
  \and
  School of Physics, University of Sydney, Sydney, NSW 2006, Australia
}

\date{Received date/ Accepted date}

\abstract
{Bisectors of strong, single spectral lines, usually the \fe\ line, have traditionally been used to examine
the velocity fields in stellar atmospheres. This requires high S/N
often achieved by summing many individual spectra.} 
{We investigate whether bisectors derived from cross-correlation functions (CCF) of single-exposure spectra can be used
to provide information on stellar atmospheres, and whether they can be used to discriminate between radial velocity 
changes caused by planets, magnetic activity and oscillations.} 
{Using a sample of bright stars observed with the HARPS spectrograph, 
we examine the shapes of the bisectors of individual strong spectral lines in summed spectra, comparing with 
similar studies in the literature.
Moreover, we examine four different quantitative CCF bisector measures
for correlations with radial velocity and stellar parameters.} 
{We show that CCF bisector measures can be used 
for quantitative analysis, employing both the absolute values and the variations. From absolute values,
$\log g$ and absolute magnitude can be approximated, and from the correlations with radial velocity one can
distinguish between magnetic activity, oscillations and orbiting planets as the probable cause of radial
velocity variations. 
We confirm that different isolated spectral lines show different bisector shapes, even between lines of the same element,
calling for caution in trying to derive global stellar properties from the bisector of a CCF. 
For the active star HR 1362 we suggest from the bisector shape an extra photospheric heating 
caused by the chromosphere of several hundred degrees.  
We confirm the fill-in of spectral lines of the Sun taken on the daylight sky
caused by Rayleigh-Brillouin and aerosol scattering, 
and we show for the first time that the fill-in has an asymmetric component.} 
{} 

\maketitle

\section{Introduction}

The analysis of high-resolution stellar spectra can reveal a wealth of information about the physical
conditions in the atmosphere. Immediate information on effective temperature, gravity and elemental abundances
comes from the analysis of line depths and equivalent widths (EW), while information on the granulation and its 
cause, the convection, is harder to interpret in terms of line profile variations. These variations
are often described in terms of the line bisector.
The physical interpretation of the line asymmetries giving rise to the different bisector shapes has been
discussed by \citet{gray2005} from an observational point of view, while theoretical modelling
has been done by \citet{asplund+2000} with good results for the Sun.

The bisector has diagnostic power for a number of stellar parameters: The effective temperature
and luminosity can be read from the shape of the bisector (the classical ``C'' shape) and from
the height of the blue-most point on the bisector, respectively \citep{gray2005}. 
Changes in the bisector shape may have several different causes:
Variations at a very low level can be introduced by reflected light from an orbiting planet \citep{hatzes+1998}.
More importantly, the
bisector is heavily affected by photospheric magnetic fields,  
and as such it has been used to distinguish radial velocity (RV) changes caused by 
planets from changes in the photospheric velocity fields induced by magnetic activity
\citep[e.g.][]{queloz+2001}. The bisector has been used to infer the presence of unseen stellar
companions, both physically connected companions as well as 
unrelated objects along the line of sight \citep[e.g.][respectively]{santos+jvc2002,torres+2005}.  
Classical stellar oscillations also introduce spectral line asymmetries that will
be visible in the bisector, but rather than the traditional bisector, other measures have often been
used for oscillation studies \citep[e.g.][]{baldry+bedding2000,dall+frandsen2002}. 

High precision single-line bisector studies require very high S/N ratios, 
which for bright stars can be achieved with a single exposure.  
For fainter stars, or if the instrument does not offer 
sufficiently high S/N in one exposure, the high S/N is often achieved by summation
of several tens to hundreds of individual spectra. 
If the summation is done with
spectra taken over a long period of time, it may introduce errors due to developments in the atmosphere
(e.g. oscillations or activity) and due to errors on the wavelength 
calibration of individual spectra. Even small wavelength offsets will have an impact on the 
summed spectral line and its bisector.   This is also true for the case where bisectors are calculated 
for many different lines in a single exposure and then averaged, since the absolute positions of
the individual bisectors are generally not known with very high precision.

The recent advent of a new generation of ultra-stable high-resolution spectrographs
has caused a revolution in the detection of extra-solar planets in the past few years.
Since the first detection of an extra-solar planet around a solar-like star
\citep[\object{51 Peg};][]{mayor+queloz1995}, the art of detecting planets by measuring RV 
displacements has been steadily refined. 
The methods used are the self-calibration with an iodine cell and the simultaneous ThAr calibration method. While the former
is self-calibrated due to the passing of the stellar light through iodine vapor, the second relies on the recording
of a ThAr calibration spectrum alongside the stellar spectrum.  The simultaneous ThAr method has the advantage
that the stellar spectrum is undisturbed, while the iodine-cell introduces hundreds of I$_2$ absorption lines in
the spectrum.  On the other hand, the ThAr method imposes much stricter requirements on the mechanical stability
of the spectrograph. 
The current state-of-the-art instrument for accurate RV measurements using the simultaneous ThAr method is beyond doubt
the ultra-stable spectrograph HARPS \citep{harps2003}, installed at the 3.6m telescope
at the La Silla site of the European Southern Observatory, which can achieve
precision better than 1~m\,s$^{-1}$ per exposure \citep{harps2004}. 
The high precision of HARPS comes from the high internal mechanical, temperature and pressure
stability, combined with the use of the simultaneous recording of a ThAr calibration lamp spectrum
alongside the stellar spectrum. In the reduction process, the calibrated spectrum is cross-correlated
with an appropriate stellar mask, matched to the spectral type of the target star.  The resulting 
cross-correlation function (CCF)
can be thought of as a ``mean'' spectral line, and as such, it has inherently high S/N. 
Iodine cell spectra have also been used for CCF bisector analysis by \citet{martinez+2005}
after proper removal of the iodine lines. However, this implies an extra reduction step, and in what
follows we will only consider CCF's obtained with the simultaneous ThAr method.
Due to the high spectral stability, calibration 
errors are negligible, and errors due to changes with time are eliminated since the CCF is constructed
from one single exposure, allowing better time resolution to study variability phenomena.  
Furthermore, in contrast to the classical bisectors, an absolute
velocity position of the HARPS CCF bisector is known to very high precision.  
On the other hand, combining many different spectral lines, spanning a wide range in
wavelength and physical conditions may wash out any physical information, and it does not discriminate 
between blended and unblended lines. Moreover,
the particular mask used in constructing the CCF will have an effect on the shape of the bisector.

In this paper we will investigate the bisectors measured with HARPS on late-type stars.  Especially, we will
investigate whether the CCF can be used as a meaningful probe of stellar atmospheric physics instead
of the classical bisectors of individual spectral lines normally used.   In order to do this we compare the 
classical bisector
found from the sum of tens to hundreds of individual spectra with the CCF bisector for a small grid of solar-like
stars, both active and non-active.

\section{Observations and data reduction}

The HARPS pipeline (Data Reduction Software; DRS) gives the reduced 1D spectrum plus a file containing
the computed CCF for all 72 orders, in addition to the mean CCF, with the computed
RV included as a  header keyword.

In order to compare the CCF bisector with the classical bisector, we have combined
a large number of 1D spectra for each target star (Table~\ref{tab:obs}) 
in order to reach as high S/N as possible. The wavelength calibration and stability 
of HARPS are good enough that negligible errors are introduced due to the adding of spectra taken on different dates.
The observations were obtained during long asteroseismology runs,
during planet searches and during technical tests of the guiding system.  The solar spectra
were taken on the daytime sky for calibration purposes during one of the
asteroseismology programmes (Kjeldsen et al., in preparation).

Each individual spectrum was first shifted to the rest wavelength using
the calculated RV and the IRAF task {\tt dopcor}, and then combined using {\tt scombine}.

\begin{table}
\caption{The target stars. $N_\mathrm{sp}$ is the number of 1D spectra
combined. The S/N was estimated from several  continuum regions in the 6200--6600~\AA\ range of the combined spectra.
Atmospheric parameters are from \citet{santos+2005} (HR 98, 6585, 7665),
\citet{allende+2004} (Sun, HR 509, 2261, 5933, 7602), 
\citet{dall+2005} (HR 1362) and this work (HR 1326, see Sect.~\ref{alfhor}).}\label{tab:obs}
\centering
\begin{tabular}{rllrrrrr}\hline\hline
HR    &  Name      &  Spec.  &  $N_\mathrm{sp}$ &  S/N &  $T_\mathrm{eff}$  &  $\log g$  &  [Fe/H] \\
      &            &  Type   &                  &                    &   [K]              &  &   \\     \hline
      & \object{Sun}       &  G2\,V          & 628  & 1200 & 5777 & 4.44 & 0.00  \\
98    & \object{\bethyi}   &  G2\,IV         & 2766 & 1700 & 5837 & 4.00 & $-$0.08  \\
509   & \object{\taucet}   &  G8\,V          & 55   & 1500 & 5328 & 4.62 & $-$0.52 \\
1326  & \object{\alfhor}   &  K1\,III        & 137  &  800 & 4670 & 2.75 & $+$0.02  \\
1362  & \object{EK Eri}    &  G8\,III        & 9    & 1300 & 5240 & 3.55 & $+$0.09   \\
2261  & \object{\alfmen}   &  G6\,V          & 36   & 1000 & 5473 & 4.51 & $+$0.03   \\
5933  & \object{\gamser}   &  F6\,IV         & 160  & 1300 & 6246 & 4.30 & $-$0.15  \\
6585  & \object{\muara}    &  G3\,V          & 275  & 1100 & 5806 & 4.28 & $+$0.32 \\   
7602  & \object{\betaql}   &  G8\,IV         & 135  & 1200 & 5106 & 3.54 & $-$0.19 \\
7665  & \object{\delpav}   &  G7\,IV         & 76   &  700 & 5614 & 4.29 & $+$0.36 \\
\hline
\end{tabular}
\end{table}

\section{Bisector definitions and tools}

The general definition of the bisector has been given by \citet{gray1988}. In short,
one finds the midpoint of the line for a number of intensity positions inside the line.
In practice, one may calculate the midpoint between all the points in the blue (red) wing 
and corresponding interpolated points in the red (blue) wing.
For our analysis we interpolate both the red and blue wing in turn, 
which results in a better sampling of the bisector.

Once the bisector has been found, several parameters can be extracted regarding the shape. Traditionally,
the so-called 7\%-span or the \emph{velocity span} is used \citep{toner+gray1988}. More recently, \citet{gray2005}
found that the height of the blue-most point on the bisector is a powerful luminosity indicator
for the cooler spectral types, 
while the temperature dependence involves the full shape of the bisector and must be probed by comparing
with tabulated values.  

\citet{queloz+2001} pioneered the use of bisectors of CCFs. To distinguish
between planet- and activity-induced RV variations, they calculated a velocity span, which they note is
essentially the inverse slope of the bisector  and analogous to the velocity span definition used for classical  
single-line bisector work. We will reserve the term \emph{velocity span} for classical bisectors, and 
we will adopt the definition of Queloz et al. of the bisector inverse slope: 
\begin{equation}
\mathrm{BIS} = v_\mathrm{t} - v_\mathrm{b},
\end{equation}
where $v_\mathrm{t}$ is the mean bisector velocity in the region between 10\% and 40\% of the line depth, and
$v_\mathrm{b}$ is the mean bisector velocity between 55\% and 90\% of the  line depth.   This definition is also similar,
but not identical, to the velocity span defined by \citet{povich+2001}. 

While the BIS is a way
to quantify the bisector, it is a purely arbitrary measurement.  In this paper we will compare
the BIS with three other bisector measurements, that are defined as follows:
(1) The \emph{bisector slope} $b_b$, which is defined as the inverse slope from a linear fit to the part of the bisector
between 25\% and 80\% of the line depth. 
The reason for this choice, is that the CCF bisectors often appear very straight in the central parts, as opposed to  
the characteristic ``C'' shape of a classical single-line bisector.
 (2) The  \emph{curvature}, defined as 
\begin{equation}
c_b = (v_3 - v_2) - (v_2 - v_1),
\end{equation}
which follows \citet{povich+2001}, except that we define the mean velocities in segments including 20-30\% ($v_1$),
40-55\% ($v_2$), and 75-100\% ($v_3$) of the
line depth.  We do not use the velocity displacement $v_b$ of Povich et al.\ because it is essentially just an
average bisector velocity position, relative to the position of the line core $\lambda_c$. However, this normalization
is purely arbitrary. A better choice would be to measure it relative to the line center i.e. the RV, which in
general does not correspond to the line core position. Instead we employ (3) the \emph{bisector bottom} $v_\mathrm{bot}$,
which is simply the average of the bottom four points on the bisector minus the RV. 
This is indeed as arbitrary a construction as
any, and can only be justified by considering that we know the absolute position of the line (the RV) to very high
precision.

 Here we must define what we mean by the absolute position of the line: While the true (in an absolute sense)
value of the RV is not known, we can measure absolute positions relative to the instrument zeropoint, which is
given with 1~\ms\ precision in this case, defined by the long-term stability of the HARPS spectrograph. This precision 
depends on the S/N, so for individual lines it will be somewhat lower than for the CCF.

\section{Bisectors of single lines}\label{singleline}

\begin{figure}
\resizebox{\hsize}{!}{
\includegraphics{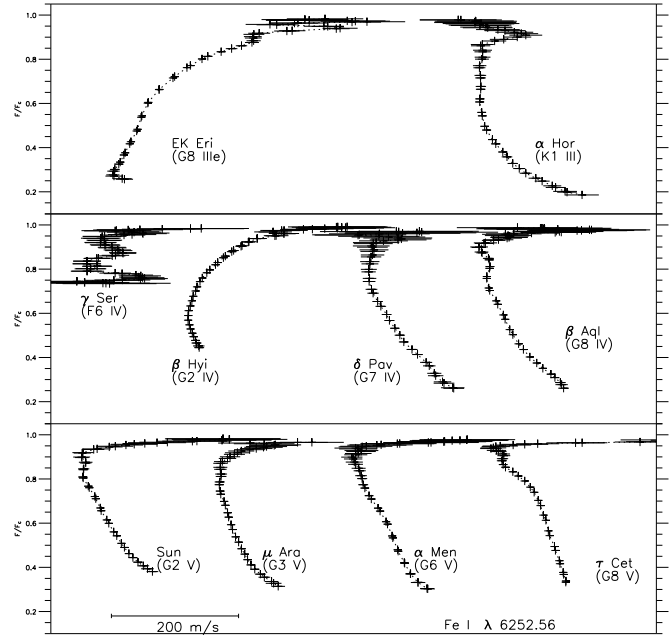}}
\caption{\label{fig:bis-all}The \ion{Fe}{i} $\lambda$\,6252 bisectors of all program stars. 
Crosses show the calculated points of the bisector with error bars on the velocity.}
\end{figure}

The bisectors of the \ion{Fe}{i} $\lambda$\,6252 line are plotted in Fig.~\ref{fig:bis-all}. 
Error bars are estimated from the measured S/N in the spectra, using the formula of \citet{gray1988}.
This is the only line 
used by \citet{gray2005}, and comparing the stars we have in common (\betaql, \taucet\ and the Sun) we find very
good overall agreement.  Hence HARPS can --- not surprisingly --- be used as a classical high-resolution spectrograph
for line-profile analysis, 
although it is necessary to sum several individual HARPS spectra to reach the high S/N.

Although in an overall sense, the bisector shapes can be understood in terms of simple temperature and luminosity
effects, there are aspects of the bisectors in Fig.~\ref{fig:bis-all} that call for attention, in particular the peculiar
appearance of the \gamser\ bisector, and the shape of the EK~Eri bisector.

As can be seen, the \ion{Fe}{i}\,6252 line is quite weak in \gamser. Moreover, the lines are broadened by rotation,
which further reduces the depth of the line, but on the other hand allows a better sampling of the profile. 
However, the star is close to the granulation boundary so 
no definite conclusions can be made. 

EK~Eri is a well-known over-active but slowly rotating star, exhibiting long-period photometric and spectroscopic variations
due to rotational modulation of starspots \citep{stepien1993,strassmeier+1999,dall+2005}.  
The bisector does not much resemble the ``canonical'' shape of a G8 giant \citep[e.g. \object{$\eta$~Dra} from][Fig.~3]{gray2005},
but rather indicates a somewhat hotter spectral type, especially towards the line core.  Since the core is formed in
the higher atmospheric layers, this could indicate a substantial heating of the upper layers by the chromospheric
activity. Fitting the bisector to the tabulated ones of Gray, we find that the lower part is better described 
by a G5 star of luminosity class III or II (no perfect fit can be obtained), indicating a heating
of the upper layers of the photosphere
by $\sim 200$--$900$~K.  This is however quite speculative. Since the star is known to have spots, the bisector
shape may be reflecting the contributions of different velocity fields in the spots. It could also simply have
a misclassified spectral type.

We will now proceed to an investigation of other potentially useful
isolated spectral lines (Table~\ref{tab:lines}).  These lines were chosen based on  VALD linelists \citep{kupka+1999}
of the solar spectrum\footnote{The Vienna Atomic Line Database, {\tt http://ams.astro.univie.ac.at/vald/}},
by requiring
that they be unblended, isolated and reasonably strong. 
The bisectors of the \ion{Ca}{i}\,6499 line were found to 
be very irregular for all our sample stars, although no obvious blends could be identified, 
and they are not plotted.  The bisectors of the remaining 
lines are shown in the Figures in Appendix~\ref{ap:lines}.
\begin{table}
\caption{All individual spectral lines investigated.}\label{tab:lines}
\centering
\begin{tabular}{llll}\hline\hline
\multicolumn{2}{l}{Line} & $\chi$ [eV] & Notes \\ \hline
\ion{Fe}{i} & 6151.62  & 2.17 & Fig.~\ref{fig:bis6151} \\
\ion{Fe}{i} & 6252.56  & 2.40 & Fig.~\ref{fig:bis-all}, (1) \\
\ion{Ca}{i} & 6499.65  & 2.52  & (2)  \\
\ion{Ni}{i} & 6643.63  & 1.67 & Fig.~\ref{fig:bis6643}  \\
\ion{Fe}{i} & 6750.15  & 2.42 & Fig.~\ref{fig:bis6750} \\
\ion{Ni}{i} & 6767.76  & 1.82 & Fig.~\ref{fig:bis6767} \\
\hline
\multicolumn{4}{l}{\tiny (1): The ``standard'' bisector line. (2): Strong blending evident in bisector, not plotted.} \\
\end{tabular}
\end{table}

The \ion{Fe}{i} $\lambda$\,6151.62 line is very well defined, but not very deep in all our stars. 
The line may be influenced by weak contributions of \ion{Ni}{i}~$\lambda 6151.67$. 
The closest major line however, is more than $0.5$\,\AA\ away (\ion{Si}{i}~$\lambda 6152.29$).
Examining Fig.~\ref{fig:bis6151} we find no obvious systematics,
and we consider this line inappropriate for stand-alone analysis. This seems to indicate, that the presence of
even very weak blends can disturb the interpretation of the bisector, provided the lines are
sufficiently close in wavelength. 

The \ion{Fe}{i} $\lambda 6252.56$ is the standard bisector line, thought to be sufficiently well isolated and free of blends. For low
rotational broadenings, the neighboring \ion{V}{i}~$\lambda 6251.83$ line is well separated, and the only other 
weak lines that may possibly contribute are all on the order of $0.5$\,\AA\ away or further, namely \ion{Ru}{i}~$\lambda 6252.07$,
and \ion{Cr}{i}~$\lambda 6253.17$.

As mentioned, the bisectors of \ion{Ca}{i}~$\lambda 6499.65$ were all very irregular. This may be ascribed to 
blending with nearby \ion{Si}{i}~$\lambda 6499.23$ and \ion{N}{i}~$6499.51$.

The \ion{Ni}{i} $\lambda$\,6643.63 line is 
very deep and well defined for most of our stars, 
and generally shows a larger velocity span than the \fe\ line. 
It is probably affected by blending with \ion{V}{i}~$\lambda 6643.79$ and \ion{Gd}{i}~$\lambda 6643.94$ on the red side and with
\ion{Sr}{i}~$\lambda 6643.53$ and \ion{Cr}{i}~$\lambda 6643.23$ on the blue side, which may contribute to the larger velocity span,
but rendering any physical interpretations very uncertain.

For the \ion{Fe}{i} $\lambda$\,6750.15 line, 
only a few of our stars show bisectors that resemble the classical ``C'' shape (Fig.~\ref{fig:bis6750}).
The line and bisector are well defined, but the sharp redward turn in the upper part may indicate
blending with another line.  The prime candidates for blending are \ion{Fe}{i}~$\lambda 6750.58$ and 
\ion{Cu}{i}~$\lambda 6749.65$, although the latter is probably too far away.

The \ion{Ni}{i} $\lambda$\,6767.76 line show 
well defined bisectors for most stars (Fig.~\ref{fig:bis6767}). However, no obvious systematics in the shape
makes it unlikely that this line can be used to infer atmospheric parameters.  It is probably disturbed by
\ion{Fe}{i}~$\lambda 6767.71$, and possibly by \ion{Co}{i}~$\lambda 6768.17$.

In conclusion, it seems that a ``well-isolated line'' translates to a minimum wavelength separation
to potential blends of $\sim 0.5$\,\AA.

\section{CCF bisectors}
It was shown by \citet{gray2005} that the blue-most point of the classical single-line bisector is a very good indicator of
luminosity class for the late G to early K stars. If an analogous relationship exists for the CCF bisector, then it will
provide us with a luminosity estimate for stars where high-S/N spectroscopy is not feasible.

\begin{figure}
\resizebox{\hsize}{!}{
\includegraphics{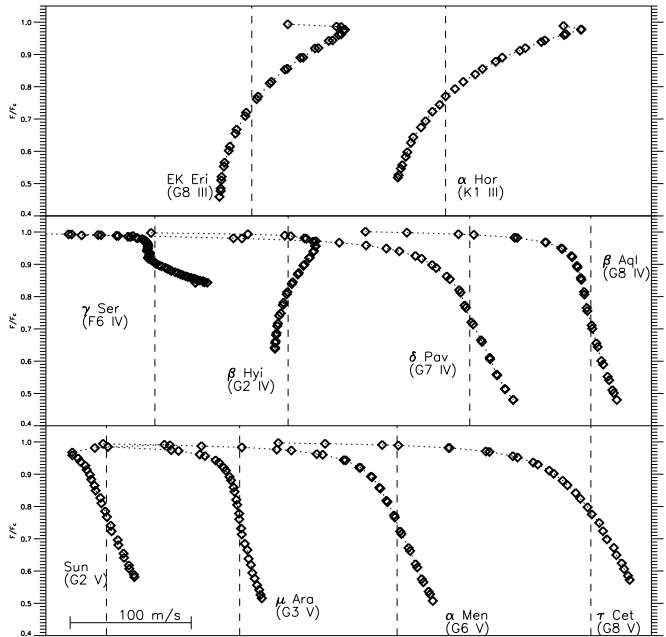}}
\caption{\label{fig:bis-ccf-all}The mean CCF bisectors of all program stars, grouped according to luminosity class.
The vertical dashed lines indicate the RV of each star.
Formal error bars are the size of the plotting symbols or smaller.}
\end{figure}

The mean CCF bisectors are plotted in Fig.~\ref{fig:bis-ccf-all}. 
Their appearances differ somewhat from the 
\ion{Fe}{i}\,6252 bisectors, which is not surprising given that they are the bisectors of a ``mean'' spectral line, 
and given the diversity of individual line bisectors evident
from Figs.~\ref{fig:bis6151}-\ref{fig:bis6767}.

\begin{figure}
\resizebox{\hsize}{!}{
\includegraphics{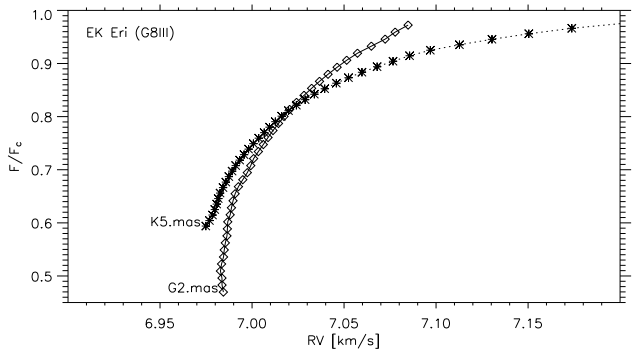}}
\caption{\label{fig:diff-bis}The CCF bisector of EK Eri computed using two different masks.}
\end{figure}

It must be stressed that one must be careful in comparing CCF bisectors of stars with different spectral types, since
the bisector will depend heavily on the mask used in computing the CCF. An example of this is shown in Fig.~\ref{fig:diff-bis}
for EK~Eri, using G2 and K5 masks. Hence, one would suspect that in general the CCF bisectors can only be used to 
compare stars of similar
spectral types, with CCFs computed using the same mask.  However, as we use a G2 mask to calculate the CCF for all stars 
except \alfhor, this precaution is of less importance in the present study.
Another concern about the masks, is whether they are constructed in a way that optimises the physical interpretation of
the CCF bisector. These masks are constructed to yield the best possible RV precision, which means excluding very weak, very
strong, and heavily blended lines. So while a given mask will include lines of many different elements, it will be free from
strong blends and hence represent a near-optimal ``average'' spectral line. 

Returning to Fig.~\ref{fig:bis-ccf-all}, there
seem to be trends in the bisector shapes, possibly with temperature, but most noticeably with luminosity and/or gravity.
\begin{figure}
\resizebox{\hsize}{!}{
\includegraphics{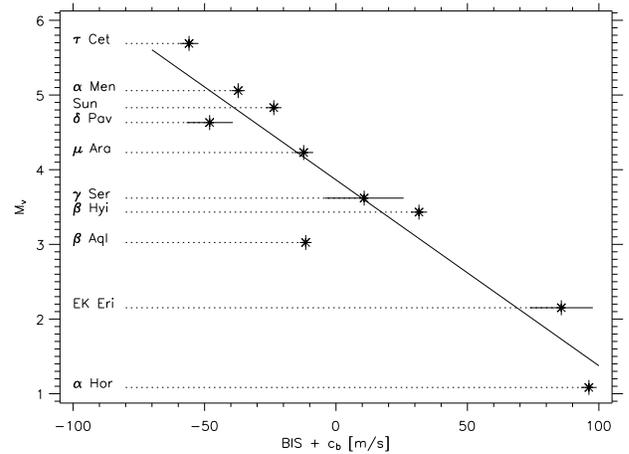}}
\caption{\label{fig:bis_vs_Mv}Absolute magnitude as a function of $\mathrm{BIS} + c_b$. }
\end{figure}
In Fig.~\ref{fig:bis_vs_Mv} we plot the absolute magnitude as a function of $\mathrm{BIS} + c_b$, which seem to exhibit
a well-defined linear relationship. A similar relationship is found plotting $\log g$ versus $\mathrm{BIS} + c_b$.
We derive the approximate relations
\begin{eqnarray}
M_v = 3.86 - 24.88 (\mathrm{BIS} + c_b) \\
\log g = 4.06 - 9.22 (\mathrm{BIS} + c_b),
\end{eqnarray}
which we must stress are only approximate, are based on very few data points, and should not be used to derive final stellar parameters.

If proven to be correct, these relationships can be used to measure the luminosity from the shape of
the average CCF bisector, in analogy with the blue-most point luminosity indicator for the classical bisector,
only that these relationships may be extendable to all the late spectral types.
However, to establish a reliable relationship, or indeed to determine whether such a relationship
exists at all, more data is needed. 
Obviously, the fact that the values of the bisector measures depends on the mask used, calls for 
extreme caution.
A more thorough analysis of a larger sample is currently in progress (Dall et al., in preparation).

\section{Variations of CCF bisectors}

In the remainder of this paper we will concentrate on the changes in the CCF bisector 
measures, rather than the absolute values of these, and 
examine how these changes may be used as diagnostic tools
when looking at only one star. In doing so, the dependence on the mask used for the CCF disappears,
as we look only at relative changes.

First, however, we must address the
expected scatter of the measures in the case where no activity or oscillations are likely to show up, using the
non-active stars with  low-amplitude oscillations, i.e. primarily \muara\ and the Sun. Some of the remaining targets
will then be addressed in turn.  

\begin{figure}
\resizebox{\hsize}{!}{
\includegraphics{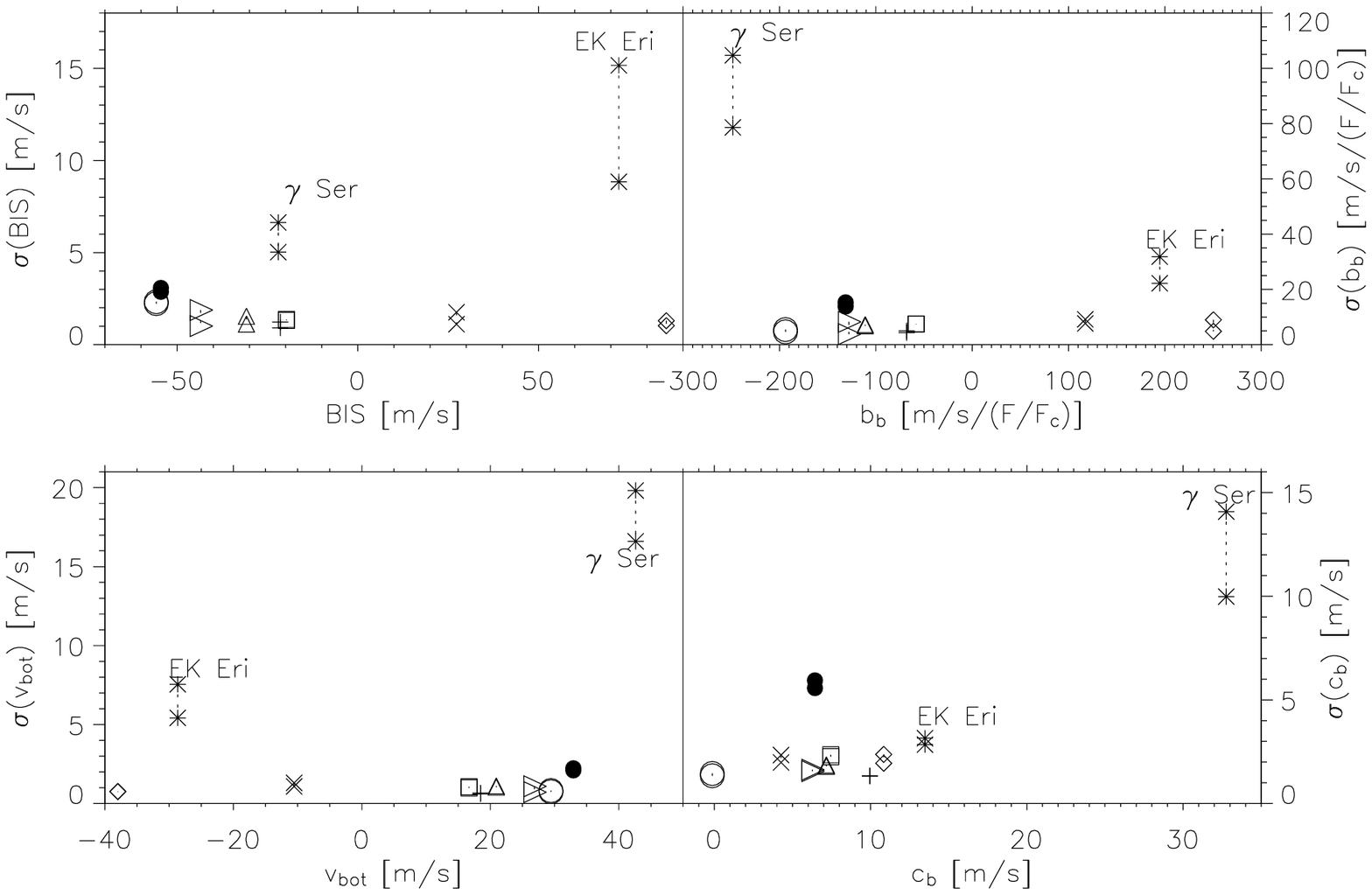}}
\caption{\label{fig:bis-scat} The scatter in all four bisector measures for all program stars. Note that
\gamser\ is close to the granulation boundary, and EK~Eri is an active star
(discussed in the text). The other symbols are for \taucet\ ($\bigcirc$), Sun ($\triangle$),
\betaql\ ($+$), \muara\ ($\Box$), \bethyi\ ($\times$), \delpav\ ($\bullet$), \alfmen\ ($\rhd$) 
and \alfhor\ ($\diamond$). The individual 1$\sigma$ RMS scatter
(upper point) and point-to-point scatter (lower point) are connected with dotted lines. 
}
\end{figure}

A general idea of the expected scatter can be gained by looking at Fig.~\ref{fig:bis-scat}, which shows
the scatter in the bisector measures for all the stars. 
Obviously, the noise will depend on the S/N
of the CCF and on the spectral type.
As can be seen, except for \gamser\ and EK~Eri, 
the scatter is very similar over the full range of spectral and luminosity classes.
In particular, the RMS spread and the 
point-to-point scatter are almost identical.

\subsection{No correlations -- \muara}
\muara\ is the host of a low-mass planet orbiting the parent star in 9.5~days \citep{santos+2004b}. Moreover,
it exhibits well-described low-amplitude solar-like  $p$-mode oscillations with individual mode amplitudes up to
0.4~\ms\ \citep{bouchy+2005,bazot+2005}.  From the ESO Science
Archive we retrieved 100 consecutive CCF observations, spanning 3.7~hours 
--- long enough to include several
oscillation cycles and short enough to avoid longer-term effects from the presence of the planet.

\begin{figure}
\resizebox{\hsize}{!}{
\includegraphics{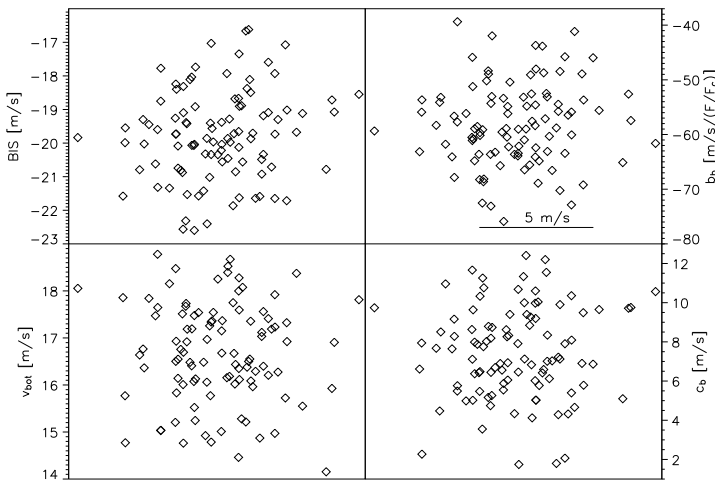}}
\caption{\label{fig:bis-ma-all}The bisector measures as functions 
of RV for \muara. The scatter is dominated by solar-like $p$-mode oscillations; see text.
Horizontal axis is RV in all plots, with the scale
indicated in the upper right panel.  }
\end{figure}

The four investigated bisector measures are shown in Fig.~\ref{fig:bis-ma-all}.
The $\sim 10$~\ms\ scatter in RV (horizontal axis) is dominated by the $p$-mode oscillations.
Note that the measured velocity is a sum of a few dozen modes, each of which 
has small ($\sim 0.4$~\ms) amplitude. These modes beat to produce the observed peak-to-peak 
velocity excursions \citep[see e.g.][Fig.~5]{bouchy+2005}. 
It is evident that no obvious correlations are present, and we will then adopt the scatter shown in these
plots to be ``typical'' of non-correlating bisector measures.

\subsection{Scattering of sunlight in the Earth atmosphere}
\begin{figure}
\resizebox{\hsize}{!}{
\includegraphics{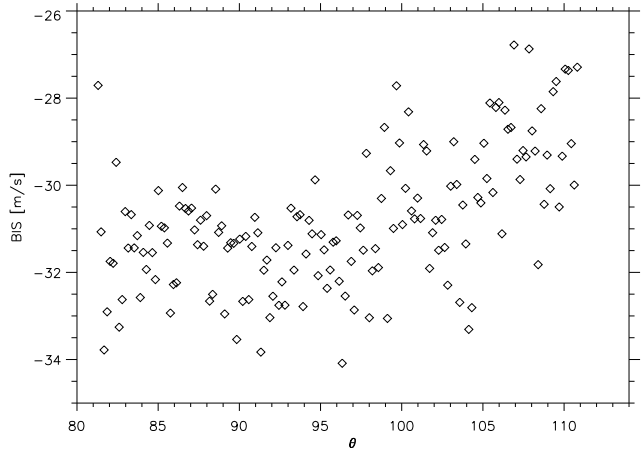}}
\caption{\label{fig:sun}The Sun: 
The bisector measure BIS as function of the angular distance to the sun $\theta$ 
for one 2.6~hour segment.  The same trend is observed in $b_b$, while no such trends can be 
seen for $v_\mathrm{bot}$ and $c_b$.
}
\end{figure}
The spectral lines in solar spectra obtained on the daytime sky are known to exhibit EW and line depth changes.
\citet{gray+2000} measured the line depth variations and found them to be symmetrical, i.e, they found 
no changes in the bisector due to asymmetries of the lines, as a function of the angular separation $\theta$
between the telescope pointing and the position of the Sun. 
Their detection limit on the bisector variations  was 20~\ms, and they mention the possibility of smaller variations being present.
They successfully modeled the line depth variations 
in terms of aerosol and Rayleigh-Brillouin
scattering as a function of $\theta$.  

Our data, taken in directions close to opposite the sun, 
covers a span in $\theta$ of $\sim 30^\circ$, in which we observe a smooth change in
the CCF line depth on the order of 1\%, consistent with the findings of \citet{gray+2000}.
From our data it is evident 
that there is an asymmetric contribution to the atmospheric scattering, as revealed in the BIS, which varies
with $4$~\ms\ over the interval (Fig.~\ref{fig:sun}).  Obviously, a better sampling of $\theta$ 
from $0^\circ$ to $180^\circ$ would be desirable, as we demonstrate that line asymmetries 
in the scattering can indeed be detected with HARPS.

\subsection{Variations due to oscillations?}
The stars \alfmen, \delpav, \bethyi\ and \alfhor\  are all expected to show solar-like $p$-mode acoustic oscillations
\citep[see][for \delpav\ and \bethyi]{kjeldsen+2005}.

Our data on \alfmen\ were taken in individual six-spectra segments of less than one hour duration, over a period
of three months.  Peak-to-peak variation is $\sim 6$~\ms, and no correlation with bisector measures can be seen.

The observations of \delpav\ span 1.8~hours, and show clear solar-like oscillations with peak-to-peak amplitudes of
$\sim 10$~\ms.  No correlation with bisector measures can be seen.  
Higher scatter of the bisector 
measures compared with the other stars is evident from Fig.~\ref{fig:bis-scat}. We have no explanation
of this fact other than the bad weather conditions (bad seeing and wind from the south) prevailing at 
the time of observation. This, and the short exposure times, could also be part of the reason for the low
S/N in the combined spectrum.

The results of the asteroseismic campaign on \bethyi\ will be published elsewhere (Kjeldsen et al., in preparation). Here it
suffice to note the peak-to-peak amplitude of the \bethyi\ oscillations of up to $\sim 10$~\ms, revealing 
weak correlations with bisector measures BIS and $b_b$.  
A Fourier analysis of the bisector measures reveals that the oscillation signal is clearly present in the BIS and $b_b$ data,
while $v_\mathrm{bot}$ and $c_b$ do not reveal any oscillation signal at the $6\sigma$ level.  This could indicate that
the upper part of the line is insensitive to the oscillations. 

\label{alfhor}
\alfhor\ is one of the most stable Hipparcos stars, with a photometric error of $< 0.3$~mmag \citep{adelman2001}.
No reliable stellar atmospheric parameters could be found in the literature, so we performed
an ATLAS9 \citep{kurucz1993} model atmosphere 
fitting using \ion{Fe}{i}/\ion{Fe}{ii} ion balance, following the procedure of \citet{dall+2005b}.
The RV curve of \alfhor\ shows a smooth variation over the duration ($\sim 2.5$~h) of the observations. The 
peak-to-peak amplitude is  $\sim 25$~\ms, maybe slightly larger as we have not covered the full
period. 
Assuming this to be due to stellar oscillations, using 
the scaling relation of \citet{kjeldsen+bedding1995} predicts a photometric amplitude of $\sim 0.3$~mmag, i.e. the oscillations
should have been barely detectable by Hipparcos.  
More importantly for our purpose, variations at this level cannot be seen in the bisector.

In summary, solar-like $p$-mode oscillations 
have a weak influence on the bisector, and are only properly revealed by  a Fourier analysis.

\subsection{Variations due to activity}
The star EK~Eri shows a clear example of activity induced bisector variations. 
The activity and associated variability of EK Eri has been studied extensively for years, and it is  
generally agreed to be a slow rotator with a fossil Ap-type magnetic field \citep{stepien1993,strassmeier+1999}.
\citet{dall+2005} found the BIS of the CCF to correlate well with the RV and
the activity index $\log R^\prime_\mathrm{HK}$, meaning that the activity displays a coherent pattern
from the photosphere to the chromosphere.

\begin{figure}
\resizebox{\hsize}{!}{
\includegraphics{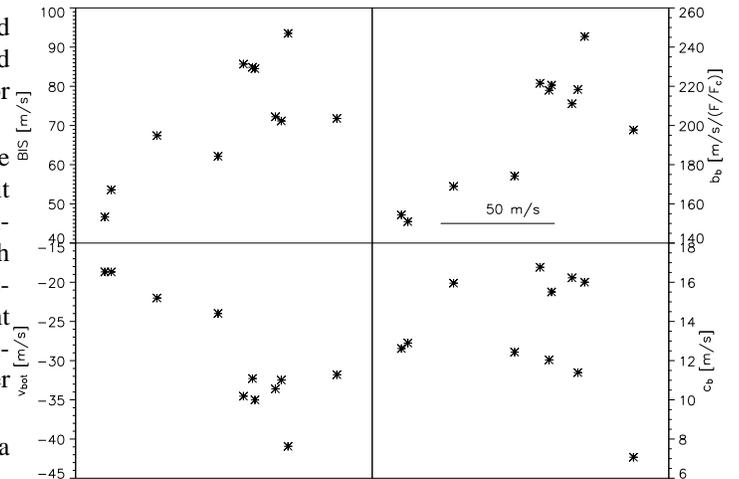}}
\caption{\label{fig:bis-ek-all}The bisector measures for EK~Eri. Horizontal axis is RV in all plots, with the scale
indicated in the upper right panel.}
\end{figure}

We examined all four bisector measures, shown in Fig.~\ref{fig:bis-ek-all}, in
order to test which is best suited to describe activity induced RV changes.
In addition to the  BIS, $b_b$ and $v_\mathrm{bot}$ show clear correlations with the RV, while
$c_b$ seem to be insensitive to the activity, which is also evident from Fig.~\ref{fig:bis-scat}.
Hence, it seems that the stability of $c_b$ is real, and that the curvature of the CCF bisector
is unaffected by activity.  Note that the slopes of the correlations are positive for BIS and $b_b$ as expected from activity
and negative for $v_\mathrm{bot}$,
and that the peak-to-peak amplitude is of the order $\sim 100$~\ms, i.e., much higher than the oscillation scatter.

\subsection{Variations due to unseen companion?}\label{bethyi}
\begin{figure}
\resizebox{\hsize}{!}{
\includegraphics{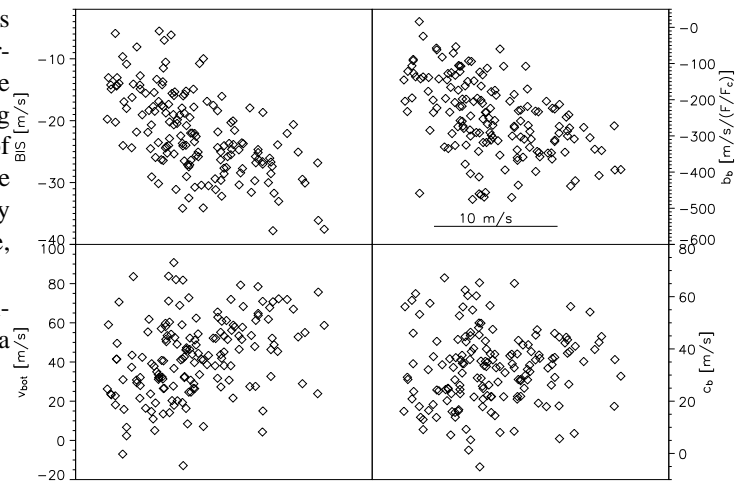}}
\caption{\label{fig:gamser}The bisector measures as functions
of RV for \gamser. Horizontal axis is RV in all plots, with the scale
indicated in the upper right panel.}
\end{figure}

For \gamser, the scatter in all the bisector measures are high (Fig.~\ref{fig:bis-scat}). A closer inspection
reveals clear correlations with BIS and $b_b$, less clear with $v_\mathrm{bot}$ and no correlation with $c_b$ (Fig.~\ref{fig:gamser}). 
Note that the slopes of the correlations are opposite to the case of the activity induced variation of EK~Eri.
This hints that the variation may be due to either a fainter orbiting star or a chance-alignment with a background star,
corresponding to the case of \object{HD 41004} \citep{santos+jvc2002} where the blending with light from a much fainter companion 
produced a negative slope of BIS versus RV. Such a correlation could
also result from chance alignment with a fainter background star, modulated by seeing variations.
Note that the possible companion discussed here is a \emph{stellar} companion, contributing
flux to the total spectrum. A planetary companion would not show up in this way. 

We hesitate to propose that the correlations could be due directly to oscillations, since the apparent amplitude 
of the RV variation is lower than for \alfhor, where no correlations were found, 
and similar to \bethyi, where only weak correlations were found.
However, the slope of the correlations are the same as for \bethyi, so this may
actually reflect oscillations, and the reason for it showing up in \gamser\ and not in \alfhor\ may lie in the 
proximity of the former to the granulation boundary. This question must be resolved by observing a larger
sample of stars.

\section{Concluding remarks}

In this paper we have examined a small sample of well-studied bright stars of different spectral types and luminosity
classes. We have measured the bisectors of isolated spectral lines using high S/N combined
spectra, and we have measured bisectors of CCFs of individual exposures, obtaining a high S/N from the 
combination of hundreds of individual lines. 

The main 
purpose of this paper --- to relate the single-line classical bisector to the single-spectrum
combined-line CCF bisector --- has been achieved: We have shown that one may use the CCF bisector in much
the same way as one would use the single-line bisector.  Moreover, we have shown that it is possible to use
the defined CCF bisector measures for quantitative analysis, employing both the absolute values and the variations.

For the single-line classical bisectors, we can point out the following:
\begin{itemize}
\item 
We confirm the general shapes of the bisectors of the \fe\ line for stars of
different spectral and luminosity classes, as published by \citet{gray2005},
using sums of HARPS spectra to achieve high S/N.  We did not find other spectral 
lines with similar well-behaved bisector patterns with spectral type. We find that to be well isolated, a line must not
have any potential blends closer than $\sim 0.5$~\AA, based on the six lines we have investigated.

\item 
Based on the bisector shape of the \fe\ line in the spectrum of EK~Eri, we  suggest that this may point to an extra heating
of the upper photosphere of several hundreds degrees, presumably supplied by the active chromosphere.
However, the peculiar bisector shape may also be due to photospheric spots, or to inaccurate spectral classification.
\end{itemize}

For the CCF bisectors, we can point out the following:
\begin{itemize}
\item 
We find that the CCF bisector could be used to approximate the luminosity and $\log g$ using the 
absolute values of the bisector measures
BIS, $b_b$, $v_\mathrm{bot}$, and $c_b$, and we derive approximate relationships.
It is slightly surprising that a luminosity relationship can be recovered in the CCF bisector, since
it includes many different spectral lines with different individual bisector shapes, not only between lines of different elements, but
between lines of the same element too (Fig.~\ref{fig:bis-all} and the Appendix). An added advantage of the CCF 
bisector in this case, is that the relationship seems to apply to all the spectral types of our sample,
and is not dependent on the existence of a blue-most point, which disappears for the hotter stars. Naturally, 
the validity and accuracy of such relationships will be the subject of future study of a larger sample of stars.

\item 
We find that normal $p$-mode solar-like oscillations do not reveal themselves in the bisectors
at a significant level (for peak-to-peak RV amplitudes $< 25$~\ms). 
However, there may be a dependence of the bisector response to oscillations on the spectral type or luminosity class.
For \gamser\ we do find correlations between RV and the bisector measures BIS and $b_b$,
even though the oscillations are at a very low level. We speculate that this could be due to an unseen stellar companion,
either a physical companion or a chance line-of-sight alignment.  Another possibility is that \gamser\ is too
close to the granulation boundary to allow any meaningful interpretation of its bisectors.

\item
For EK Eri, the only active star of our sample, we find the RV correlated with 
all bisector measures except the curvature $c_b$.
Note that the slopes of the correlations
for \gamser\ and EK~Eri are reversed, giving an indication of different mechanisms being involved.

\end{itemize}

Finally, we confirm the effects of aerosol and Rayleigh-Brillouin scattering on solar spectra taken on the
daylight sky. Moreover, we show for the first time that the bisectors and hence the lines change shape, thus
there is an asymmetric component of the scattering.

\begin{acknowledgements}
This research has made use of the SIMBAD database,
operated at CDS, Strasbourg, France.
Support from Funda\c{c}\~ao para a Ci\^encia e a Tecnologia (Portugal)
to N.C.S. in the form of a scholarship 
(reference SFRH/BPD/8116/2002) and a grant (reference
POCI/CTE-AST/56453/2004) is gratefully acknowledged.
Support from The Danish Natural Science Research Council
is gratefully acknowledged.
\end{acknowledgements}

\bibliographystyle{bibtex/aa}
\bibliography{../refs.active_stars}


\clearpage
\appendix

\section{Bisectors of additional lines}\label{ap:lines}

\begin{figure}
\resizebox{\hsize}{!}{
\includegraphics{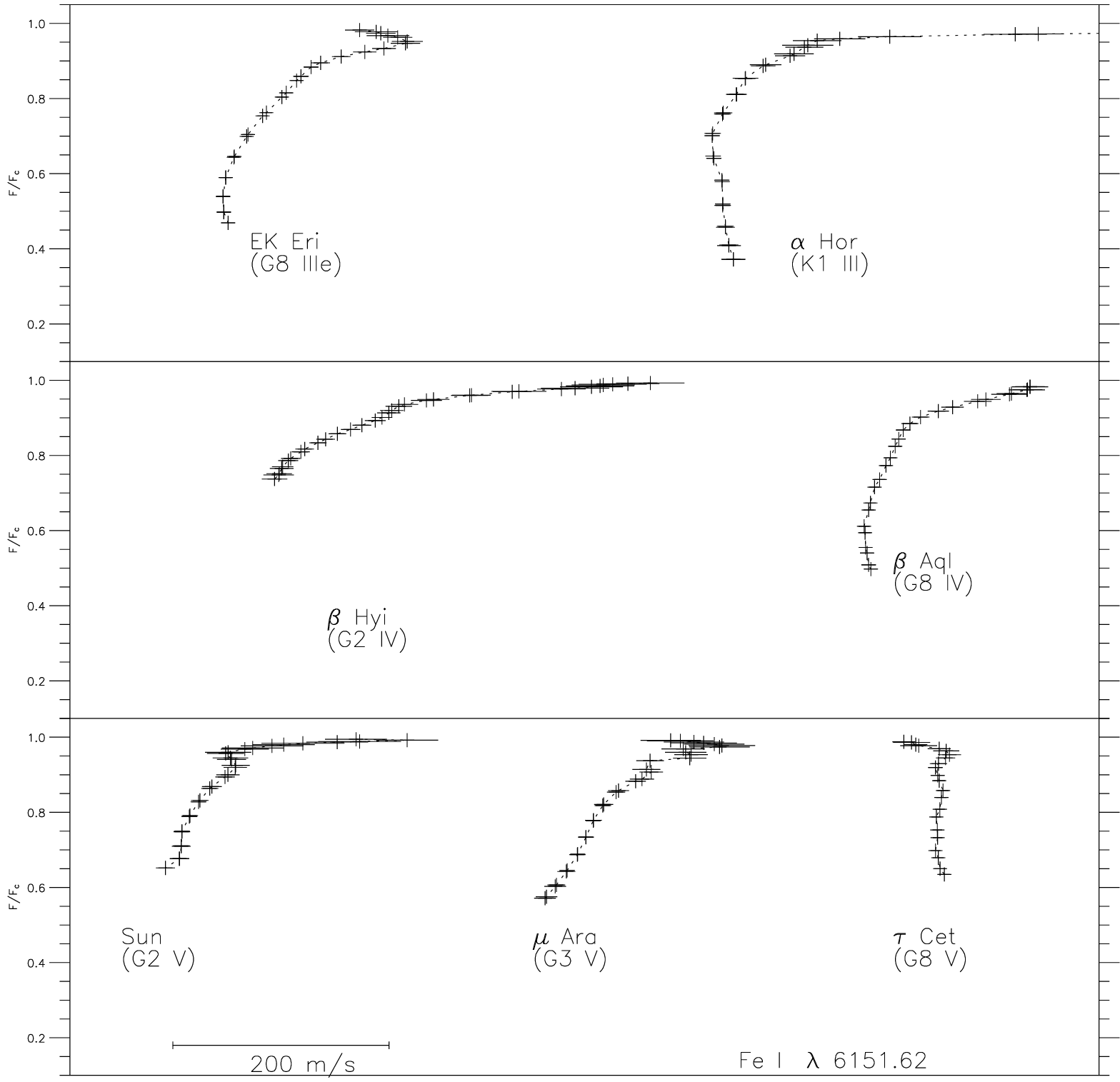}}
\caption{\label{fig:bis6151}  }
\end{figure}

\begin{figure}
\resizebox{\hsize}{!}{
\includegraphics{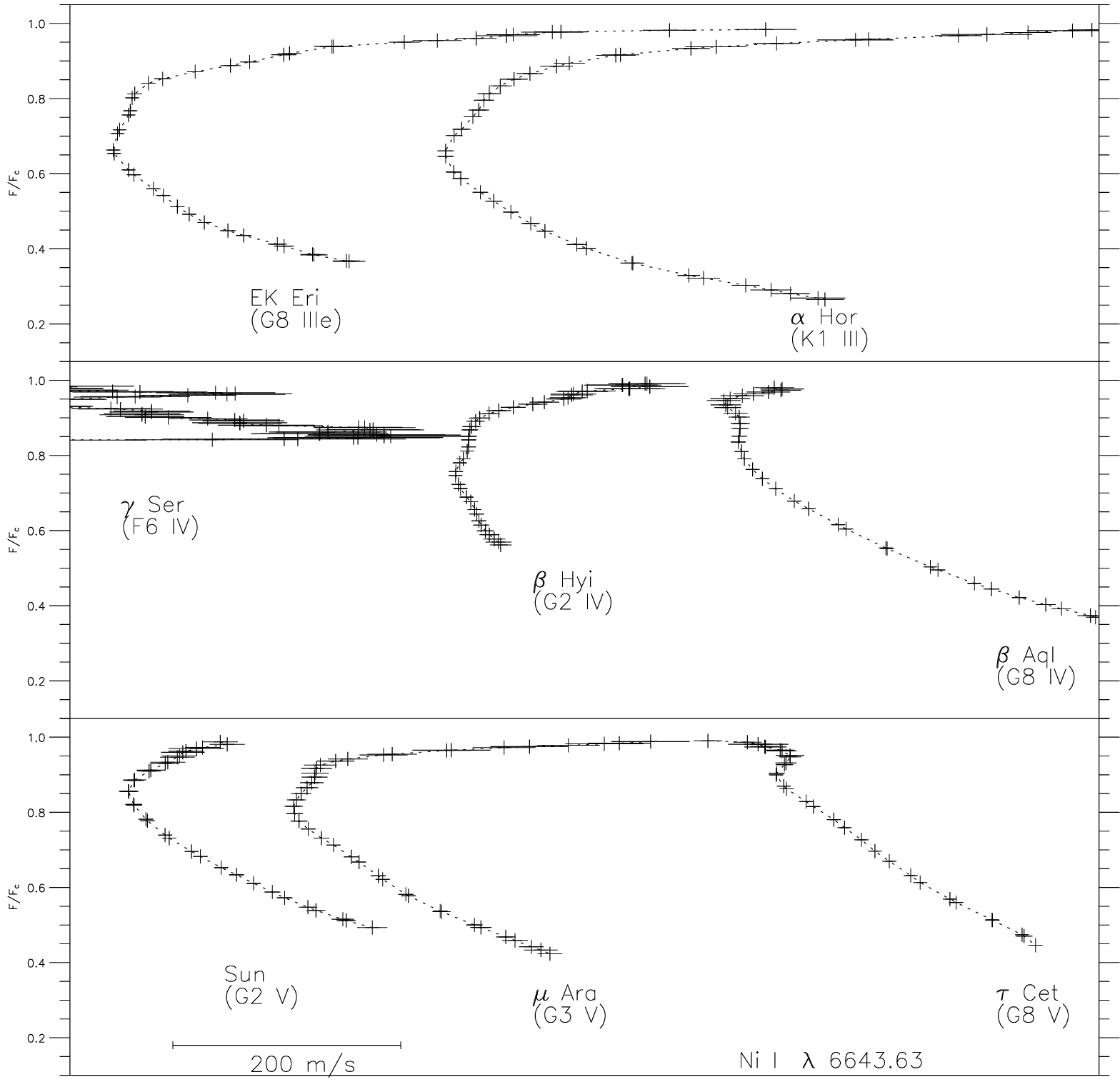}}
\caption{\label{fig:bis6643}  }
\end{figure}

\begin{figure}
\resizebox{\hsize}{!}{
\includegraphics{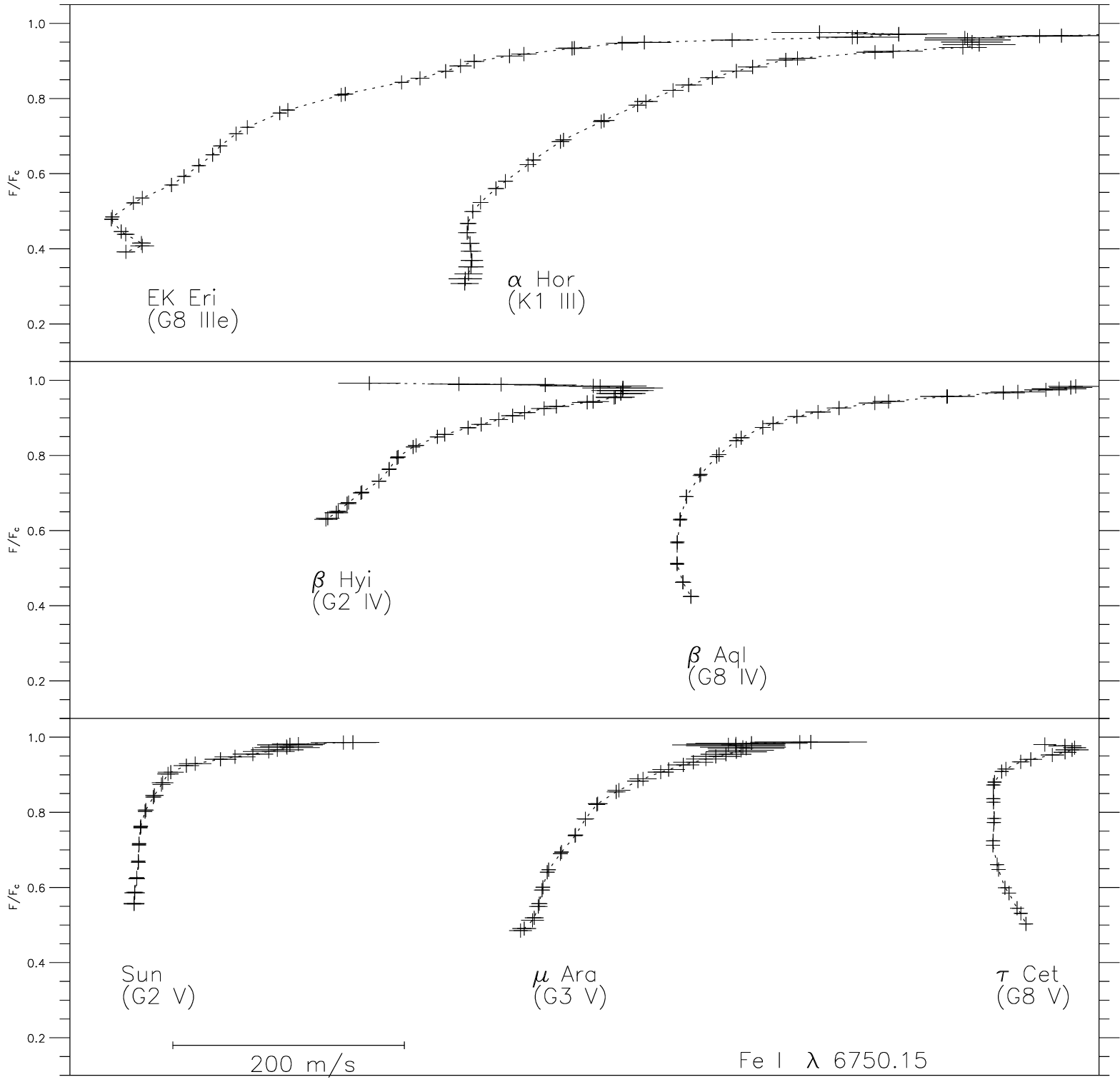}}
\caption{\label{fig:bis6750}  }
\end{figure}

\begin{figure}
\resizebox{\hsize}{!}{
\includegraphics{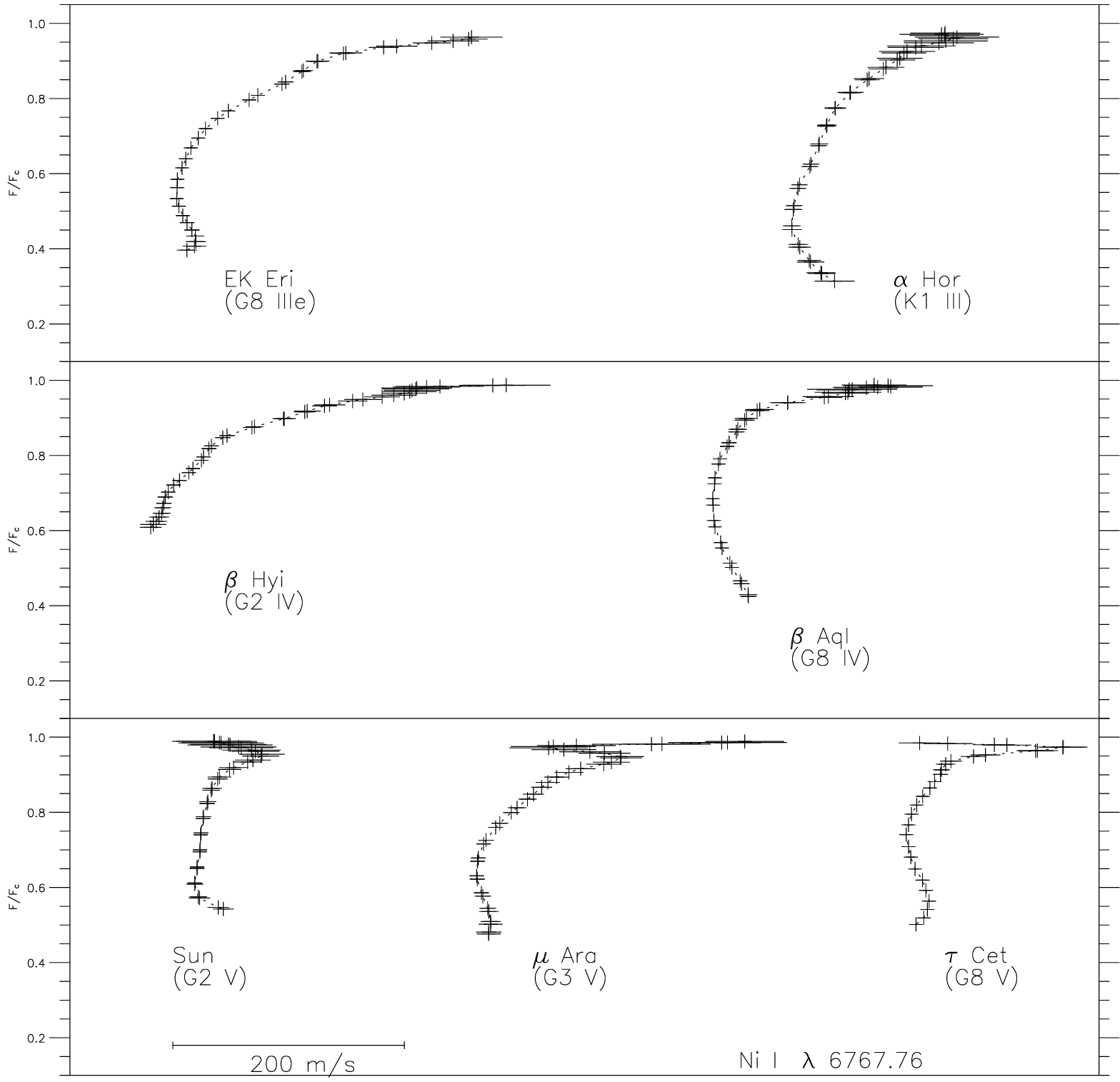}}
\caption{\label{fig:bis6767}  }
\end{figure}

\end{document}